\documentclass[aps,twocolumn,prd,amsmath,nofootinbib,preprintnumbers,superscriptaddress]{revtex4-1}
\usepackage{graphicx,slashed,hyperref,color}
\usepackage{amssymb}
\usepackage[normalem]{ulem}
\usepackage[utf8]{inputenc}
\hypersetup{
   bookmarks=true,         
   unicode=true,          
   pdftoolbar=true,        
   pdfmenubar=true,        
   pdffitwindow=false,     
   pdfstartview={FitH},    
   pdftitle={My title},    
   pdfauthor={Author},     
   pdfsubject={Subject},   
   pdfcreator={Creator},   
   pdfproducer={Producer}, 
   pdfkeywords={keyword1} {key2} {key3}, 
   pdfnewwindow=true,      
   colorlinks=true,       
   linkcolor=blue,          
   citecolor=magenta,        
   filecolor=magenta,      
   urlcolor=cyan           
}

\newcommand{\be}{\begin{equation}}
\newcommand{\ee}{\end{equation}}
\def\bsp#1\esp{\begin{split}#1\end{split}}
\newcommand\sss{\scriptscriptstyle}

\newcommand{\amc}{{\sc Mad\-Graph5\textunderscore}a{\sc MC@NLO}}
\newcommand{\fr}{{\sc Feyn\-Rules}}
\newcommand{\nloct}{{\sc NLOCT}}

\begin{document}
\title{Investigating the jet activity accompanying the production at the LHC of
a massive scalar particle decaying into photons}

\author{Benjamin~Fuks}
\affiliation{Sorbonne Universit\'es, UPMC Univ.~Paris 06, UMR 7589, LPTHE, F-75005 Paris, France}
\affiliation{CNRS, UMR 7589, LPTHE, F-75005 Paris, France}
\author{Dong Woo Kang}
\affiliation{Department of Physics, Sungkyunkwan University, Suwon 440-746 Korea}
\affiliation{Dept. of Physics \& IPAP, Yonsei University, Seoul 03722 Korea}
\author{Seong Chan Park}
\affiliation{Dept. of Physics \& IPAP, Yonsei University, Seoul 03722 Korea}
\affiliation{Korea Institute for Advanced Study (KIAS), Seoul 02455 Korea}
\author{Min-Seok Seo}
\affiliation{Center for Theoretical Physics of the Universe,
Institute for Basic Science, 34051 Daejeon, Korea}

\begin{abstract}
We study the jet activity that accompanies the production by gluon fusion
of a new physics scalar particle decaying into photons at the LHC. In the
considered scenarios, both the production and decay mechanisms are governed by
loop-induced interactions involving a heavy colored state. We show that the
presence of large new physics contributions to
the inclusive diphoton invariant-mass spectrum always implies a significant
production rate of non-standard diphoton events containing extra hard
jets. We investigate the existence of possible handles that could provide a way
to obtain information on the underlying physics behind the scalar resonance, and
this in a wide mass window.
\end{abstract}

\maketitle

\section{Introduction \label{sec:introduction}}

The resonant production of a highly massive diphoton system consists of a
prediction of many theories beyond the Standard Model, in particular in the case
where the Higgs sector is non-minimal. Scrutinizing LHC proton-proton collisions
giving rise to events featuring two hard photons plays thus a key role in the
LHC experimental program, in particular as the diphoton channel is
experimentally clean and associated with a small Standard Model background.
The related Run~I ATLAS and CMS analyses have hence been cornerstones for the
Higgs boson discovery in 2012~\cite{Aad:2012tfa,Chatrchyan:2012xdj}, and the
Run~II has a great potential to observe a new massive particle decaying into two
photons for masses ranging up to a few TeV~\cite{ATLAS-CONF-2015-081,%
CMS:2015dxe, ATLAS-CONF-2016-018, CMS-PAS-EXO-16-018}. More precisely, such a
new particle should appear as a resonant bump in the diphoton invariant-mass
spectrum. Previous hints at the $2\sigma$-level for such a diphoton resonance
have spurred an intense theoretical activity over the last few months, and
different attempts have been performed in order to characterize the excess both
from the top-down and bottom-up approaches.  In the meantime,
updated LHC results have been reported and a new physics signal now turns
out to be statistically disfavored~\cite{ATLASnew, CMSnew}. In this work, we
motivate the
study of less inclusive channels in order to verify the compatibility of the
properties of  any new state that would be decaying into a diphoton
system with respect to QCD radiation.

\begin{figure}
\centering
  \includegraphics[width=.5\columnwidth]{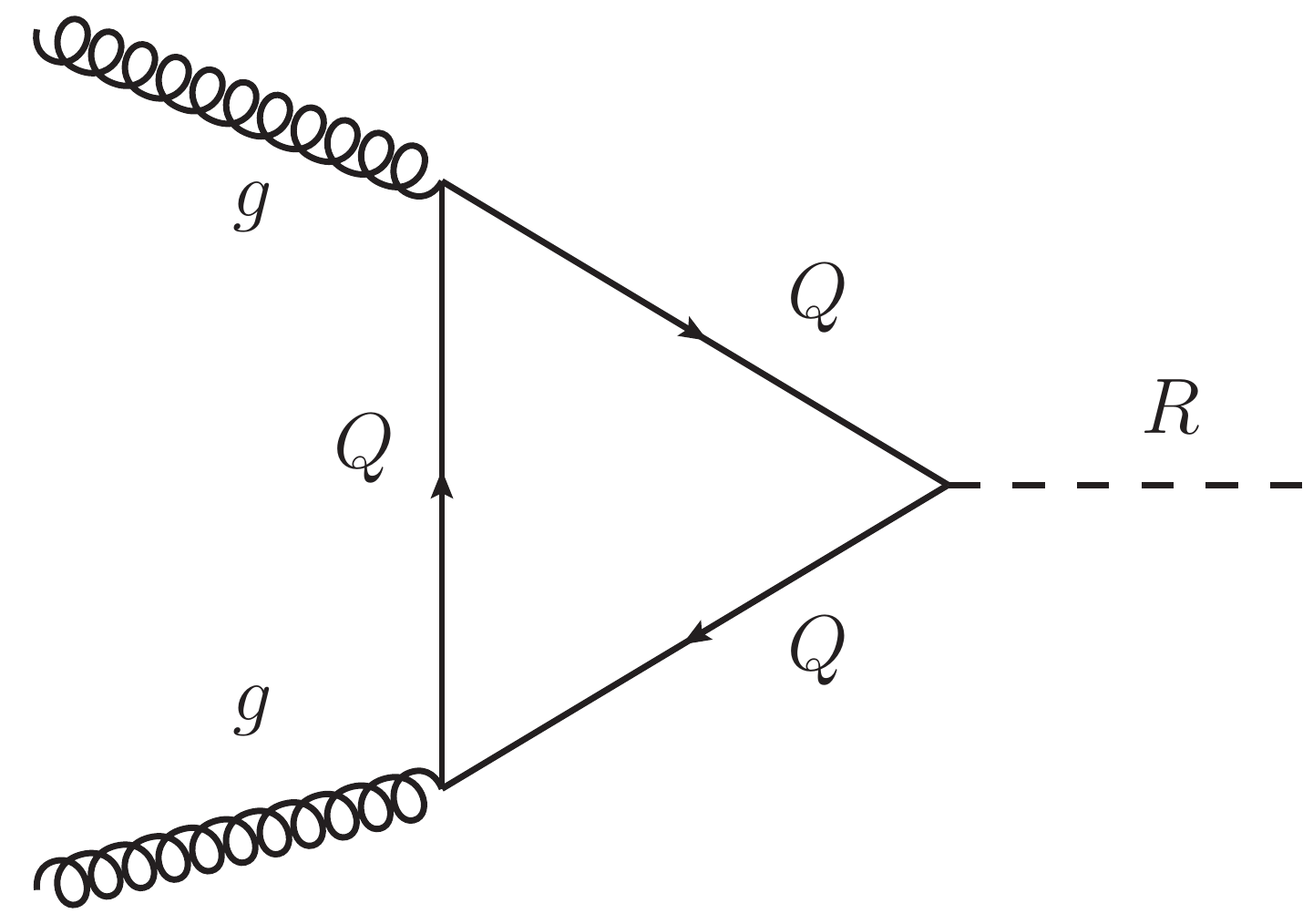}
  \caption{\small \label{fig:diagram1} Feynman diagram representing the
    loop-induced production of a scalar $R$-particle via gluon fusion when a
    heavy quark $Q$ runs into the loop.}
\end{figure}

As the Landau-Yang
theorem~\cite{Landau:1948kw,Yang:1950rg} forbids the on-shell coupling of a
massive vector boson to a photon pair and the off-shell case does not give rise
to any resonant behavior, diphoton resonance searches are usually interpreted as
limits on the existence of a scalar (with a spin $s=0$) or a tensor (with a spin
$s=2$) state.
In this paper, we focus on new physics setups featuring a massive scalar
particle that we denote by $R$ and refer to Ref.~\cite{Bernon:2016dow} for
information on the spin-two case. In order for this particle to be produced
with a
sufficiently large rate to be observed in the diphoton mode, we assume that the
couplings of the $R$ particle to the Standard Model gluons and photons are
issued from interactions with a new colored and electrically charged particle.
We investigate two simplified models where this colored particle is either a
heavy quark $Q$ or a heavy scalar quark $\tilde q$. Both scenarios yield the
loop-induced production of the $R$ state via the gluon-fusion mechanism
$gg\to R$, as depicted in Figure~\ref{fig:diagram1} for the heavy quark case,
and the $R$-decay mode into a photon pair $R\to \gamma\gamma$.

\begin{figure}
\centering
  \includegraphics[width=.95\columnwidth]{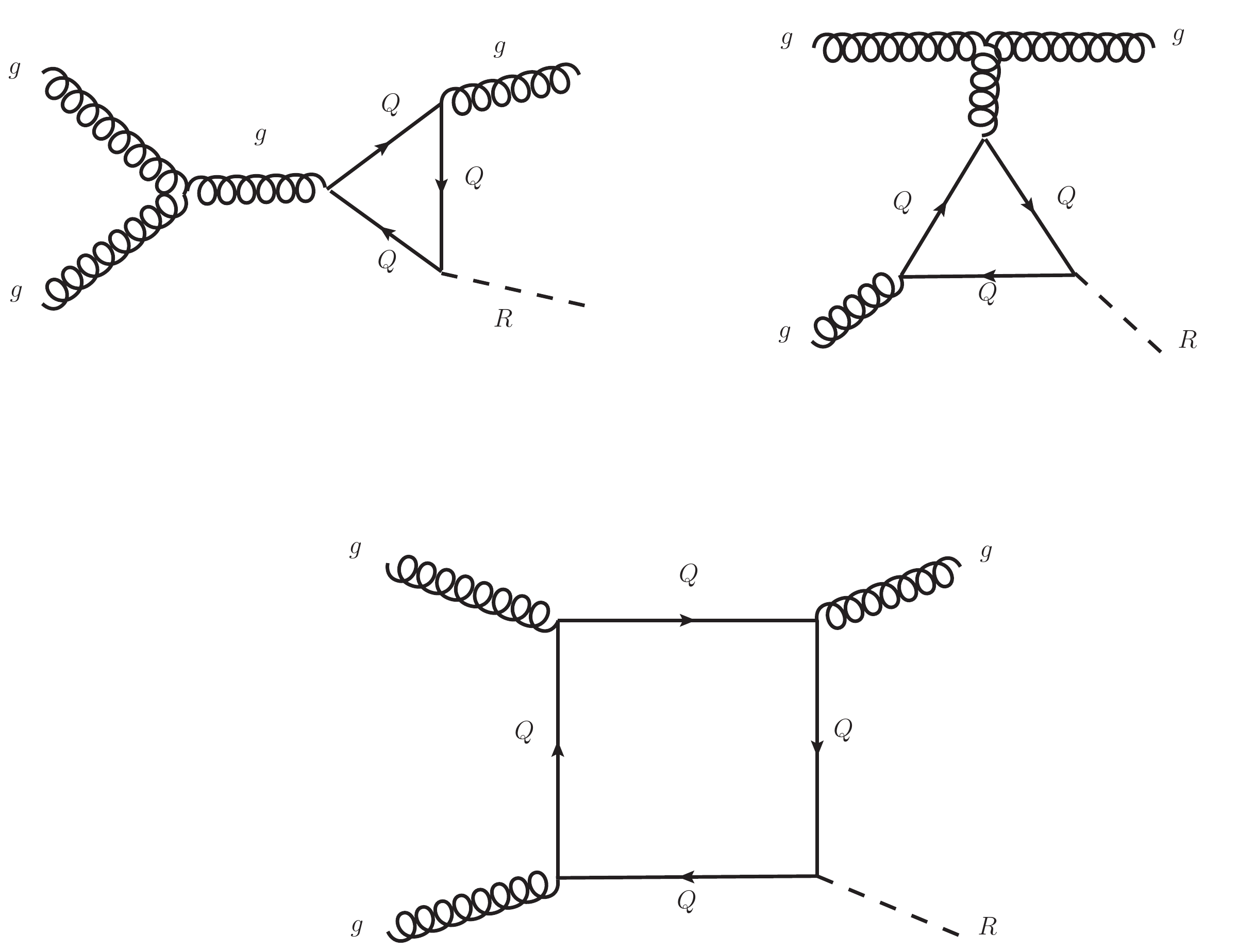}
  \caption{\small \label{fig:diagram2} Feynman diagrams representing the
    loop-induced production of a scalar particle $R$ in association with a hard
    jet, via gluon fusion and when a heavy quark $Q$ runs into the loop. The two
    diagrams in the first row are triangle ($\triangle$) diagrams and the one in
    the second row is a box ($\square$) diagram.}
\end{figure}

As shown in Figure~\ref{fig:diagram2} where we present the leading-order Feynman
diagrams corresponding to the production of the scalar particle $R$
with an additional gluon via a loop containing the heavy quark state $Q$,
the interactions under consideration also induce the associated production of
the $R$ particle with additional jets. We restrict our study to a
configuration featuring one extra jet with a transverse-momentum of at least
50--100~GeV, as the magnitude of the cross section related to
the production of the $R$ state with two or more such hard jets is estimated
to be too small to yield any statistically meaningful deviation within the
context of the current amount of recorded LHC data. There are two categories of
diagrams contributing to the $gg\to R g$ process respectively containing a
triangular loop (the so-called {\it triangle} diagrams $\triangle$ presented in
the upper panel of the figure) and a rectangular loop (the so-called {\it box}
 diagram $\square$ of the lower panel of the figure), the triangle diagrams
sharing the same vertex structure as for single $R$ production.
Since this process is loop-induced, the corresponding scattering amplitude is
ultraviolet finite. It nonetheless exhibits infrared divergences related to the
extra gluon that could be soft, collinear or both. The phase space integration can
however be safely performed thanks to the requirement of this gluon being hard,
which prevents from entering the infrared sensitive phase
space regions. In the following, we investigate how measurements of the
properties of the jets produced in association with a scalar diphoton resonance
can provide new handles to probe the underlying physics. Our study complements
recent works in which
the jet multiplicity spectrum associated with the production of the $R$ particle
with a mass fixed to 750~GeV has been probed to get information on the
production mechanism from which the diphoton signal
originates~\cite{Bernon:2016dow,Harland-Lang:2016vzm,Ebert:2016idf} or in which
effective field theory limits are studied~\cite{delaPuente:2016pyh}.

While the latter work mostly focuses on
non-renormalizable dimension-five operators describing the new physics, we in
contrast consider a theoretical framework containing only renormalizable
four-dimensional operators. By performing exact one-loop calculations, we probe
both the structure of the
loop-diagrams giving rise to a diphoton signal with and without extra hard jets
and the structure
of the couplings of the $R$ particle to the Standard Model.

In the next section (Section~\ref{sec:model}), we describe the theoretical setup
employed for our analysis and then present our phenomenological results in
Section~\ref{sec:LHCpheno}. We conclude and summarize our findings in
Section~\ref{sec:conclusion}.

\section{Theoretical framework\label{sec:model}}

In our study, we assume the existence of a real scalar field $R$ with a mass
$m_{\sss R}$ whose interactions with the Standard Model are mediated via the
exchange of a heavy quark $Q$ of mass $m_{\sss Q}$. Vector-like quarks are
important in many extensions of the Standard Model (\textit{e.g.} in theories
with extra dimensions~\cite{Park:2009cs, Kong:2010qd} or composite-Higgs models
with partial compositeness~\cite{Kaplan:1991dc}) and are generally considered
as lying in the
fundamental representation of the QCD gauge group ${\rm SU}(3)_c$.  Although they can possibly lie in many different representations of the electroweak symmetry group
${\rm SU}(2)_L\times {\rm U}(1)_Y$, we focus on a minimal setup where the $Q$ quark is a weak
isospin singlet with an hypercharge quantum number set to 2/3. 
In addition, we neglect any mixing with the Standard Model up-type
quark sector, so that we rely on the effective new physics Lagrangian
\be\bsp
 {\cal L}^{(1)}_{\rm NP} =&\
 i \bar Q \slashed{D} Q - m_{\sss Q} \bar Q Q
 +  \frac 12 \partial_\mu R \partial^\mu R - \frac 12 m_{\sss R}^2 R^2\\
 &\ + \hat\kappa_{\sss Q} R\ \bar Q  Q \ ,
\esp\label{eq:LVLQ}\ee
that is supplemented to the Standard Model Lagrangian ${\cal L}_{\rm SM}$. 
The interaction strength between the heavy quark and $R$ is
denoted by $\hat\kappa_{\sss Q}$, and the gauge covariant derivative is given by
\be
  D_\mu Q  = \partial_\mu Q - i g_s T_a G_\mu^a Q - i \frac23 e A_\mu Q\ ,
\ee
with $g_s$ and $e$ denoting the strong and electromagnetic coupling constants,
$G_\mu$ and $A_\mu$ the gluon and photon fields and $T_a$ the fundamental
representation matrices of $SU(3)$. Since the extra quark $Q$ does
not mix with the Standard Model quark sector, our new physics model
evades by construction all existing searches for vector-like quarks at the
LHC~\cite{Aad:2015tba,Aad:2015kqa,CMS:2016ccy,CMS:2015alb,ATLAS:2016sno}. As a
result, our simple parameterization also embeds models in which there are more
than one state connecting the $R$ scalar to the Standard Model. This
can be accounted for by a stronger $\hat\kappa_{\sss Q}$ coupling.

Alternatively, one may consider that the mediation of the new physics
interactions of the $R$ particle with the Standard Model proceeds via the
exchange of a scalar quark $\tilde q$ of mass $m_{\sss \tilde q}$.
For the sake of minimality and simplicity, the squark $\tilde q$ is
considered as a weak isospin singlet and lies in the fundamental
representation of the $SU(3)_c$ group. The
new physics sector is then described by the Lagrangian
\be\bsp
 {\cal L}_{\rm NP}^{(2)} =&\
 D_\mu\tilde q^\dag D^\mu\tilde q - m^2_{\sss \tilde q} \tilde q^\dag \tilde q
 +  \frac 12 \partial_\mu R \partial^\mu R - \frac 12 m^2_{\sss R} R^2\\
 &\ + \hat\kappa_{\sss \tilde q} R\ \tilde q^\dag \tilde q\ .
\esp\label{eq:Lsq}\ee
In the next section, we will investigate the effects related to the nature of
the particle connecting the new physics to the Standard Model sectors, and make
predictions by using either the Lagrangian ${\cal L}^{(1)}_{\rm NP} $ or the
Lagrangian ${\cal L}^{(2)}_{\rm NP} $. As for the vector-like quark
case, the squark $\tilde q$ does not singly couple to the Standard Model so that
our new physics modeling cannot be probed by typical squark searches at the LHC.

\begin{figure*}
\centering
  \includegraphics[width=.85\columnwidth]{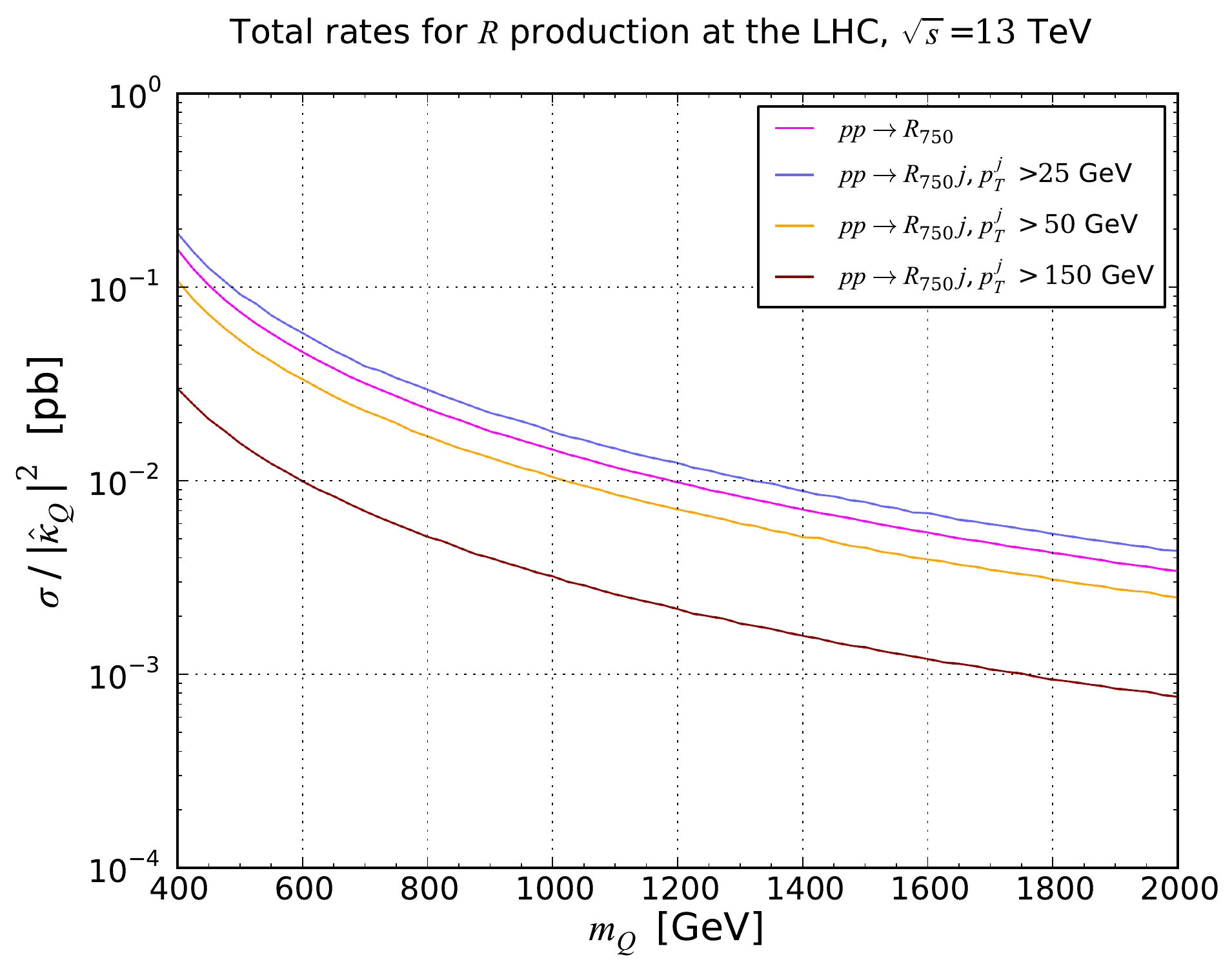}\qquad\qquad
  \includegraphics[width=.85\columnwidth]{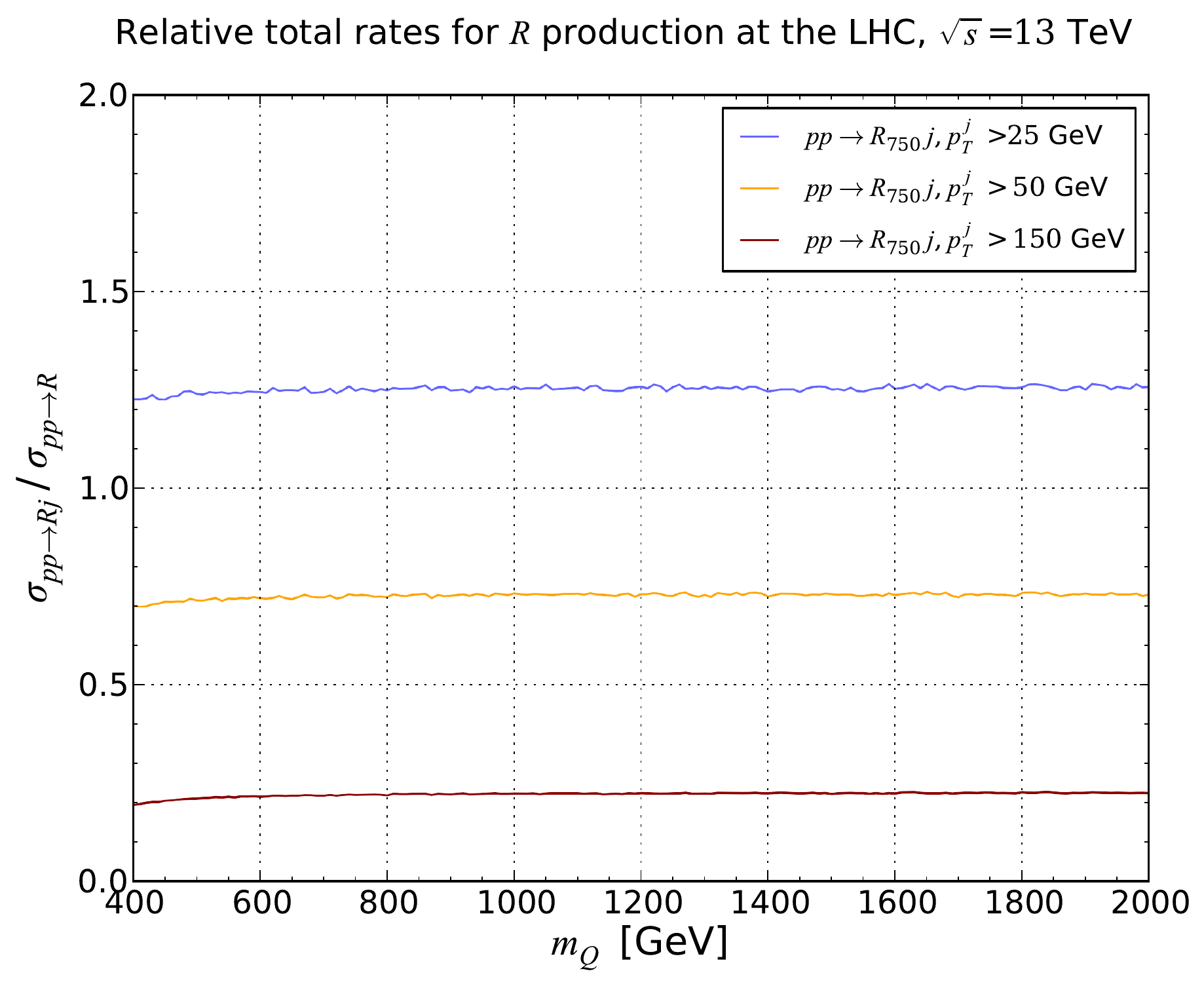}\\
  \caption{\small \label{fig:totalrates}Total production cross sections for the
  production of a 750~GeV scalar resonance, possibly in association with a jet
  whose transverse momentum is imposed to be above a specific threshold. In the
  right panel of the figure, we present the production rates relatively to the
  $p p \to R$ total production cross section. All results are presented as
  function of the mass of the vector-like quark $Q$ running into the loop
  $m_{\sss Q}$.}
\end{figure*}

In order to calculate (loop-induced) differential and total cross sections
related to processes involving an $R$ scalar in the final state, we make use of
the \amc~platform~\cite{Alwall:2014hca} whose loop-module~\cite{Hirschi:2011pa}
has been recently extended to deal with loop-induced
processes~\cite{Hirschi:2015iia}. This relies on the numerical
evaluation of loop integrals in four dimensions, which necessitates the
calculation of rational terms associated with the $\epsilon$-dimensional
components of the loop-integrals that should be normally evalulated in
\mbox{$D=4-2\epsilon$}
dimensions. These rational terms can be split into two ensembles, the first one
being connected to the loop-integral denominators ($R_1$) and the second one to
the loop-integral numerators ($R_2$). While the $R_1$ terms are universal, the
$R_2$ terms are model-dependent and process-dependent. They can however be cast
as a finite number of counterterm Feynman rules derived from
the bare Lagrangian~\cite{Ossola:2008xq}. Starting from the two
${\cal L}_{\rm NP}^{(i)}$ Lagrangians, we translate, by a joint use of the
\fr~\cite{Alloul:2013bka} and \nloct~\cite{Degrande:2014vpa} packages,
the model information into a UFO library~\cite{Degrande:2011ua} that contains all relevant $R_2$ counterterms and that is ready to be used in \amc.

\section{Phenomenological results}
\label{sec:LHCpheno}
As a first benchmark scenario, we focus on the possible characterization of the
750~GeV resonance whose hints have been recently observed by both the ATLAS and
CMS collaborations~\cite{ATLAS-CONF-2015-081,CMS:2015dxe}.
To this aim, it is useful to study its production together with a possible additional hard jet.

In Figure~\ref{fig:totalrates}, we consider a model where the coupling of the
$R$ scalar with the Standard Model occurs via heavy $Q$-quark exchanges (the
model described by the ${\cal L}_{\rm NP}^{(1)}$ Lagrangian of
Eq.~\eqref{eq:LVLQ}), and we
evaluate total cross sections for $R$ production possibly in association with a
jet whose transverse momentum $p_T$ is constrained to be above some threshold
$p_T^j$. In our calculations, the loop-induced hard matrix elements have
been convoluted with the next-to-leading order set of NNPDF~3.0 parton
distribution functions~\cite{Ball:2014uwa} accessed via the LHAPDF~6
library~\cite{Buckley:2014ana}, and we have fixed both the factorization and
renormalization scales to half the transverse mass of all final state
particles. The results are presented with the $\hat\kappa_{\sss Q}$ dependence
of the cross sections factorized out, since the latter parameter can always be
tuned so that the rate for $pp\to R$ accomodates any diphoton excess that would
be observed, and as a function of the mass of the heavy quark $Q$ that runs into
the loops. Focusing, for the sake of the example, on a benchmark scenario in
which $m_{\sss R}=750$~GeV and that corresponds to the Run~II ATLAS and CMS
past observations, vector-like quark masses ranging up to at most 2~TeV could
accomodate a signal cross section of the order of 10--20~fb and simultaneously
forbid the $\hat\kappa_{\sss Q}$ value to be such that perturbation theory
breaks down ($\hat\kappa_{\sss Q} \lesssim 10$).

\begin{figure*}
\centering
  \includegraphics[width=.85\columnwidth]{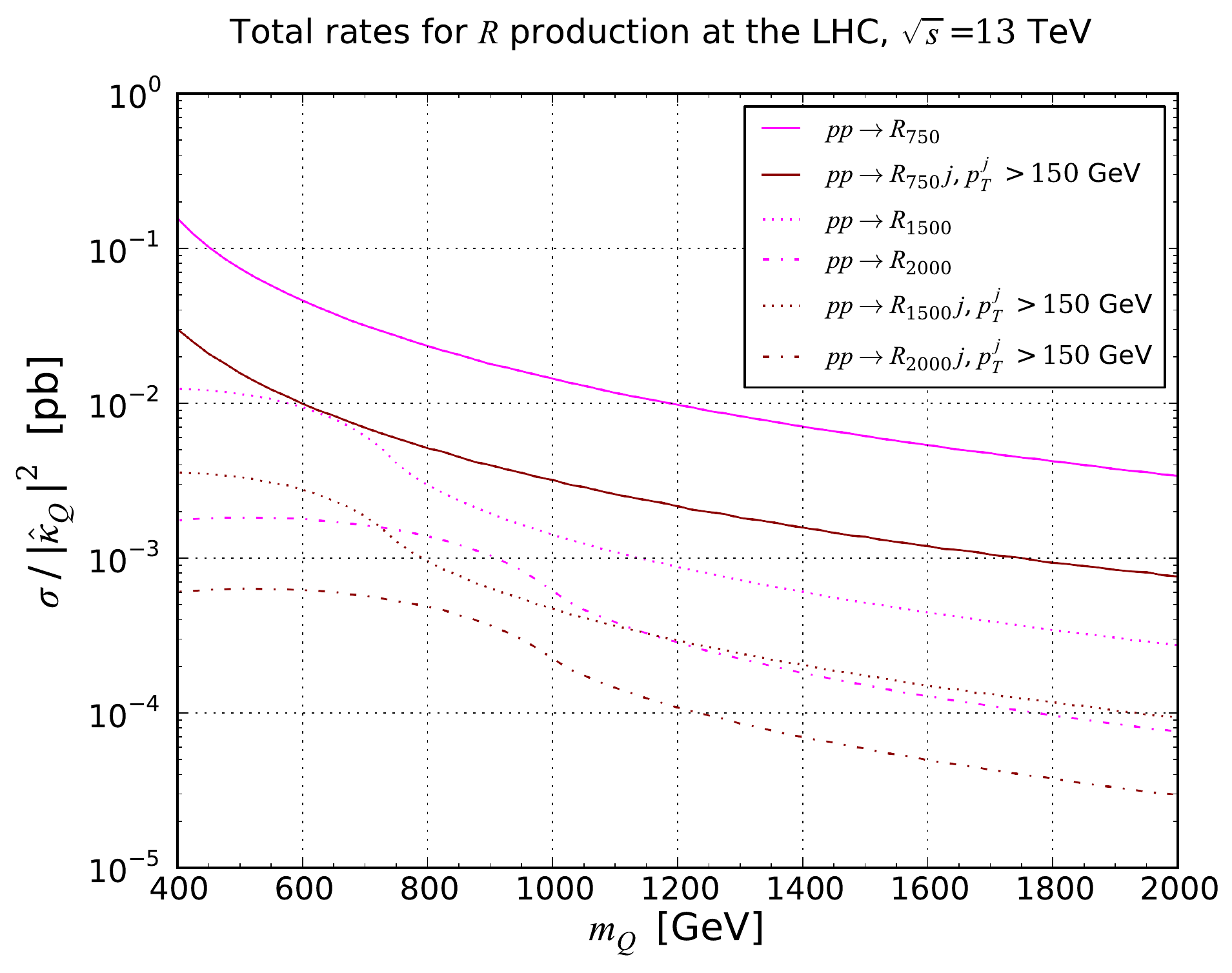}\qquad\qquad
  \includegraphics[width=.85\columnwidth]{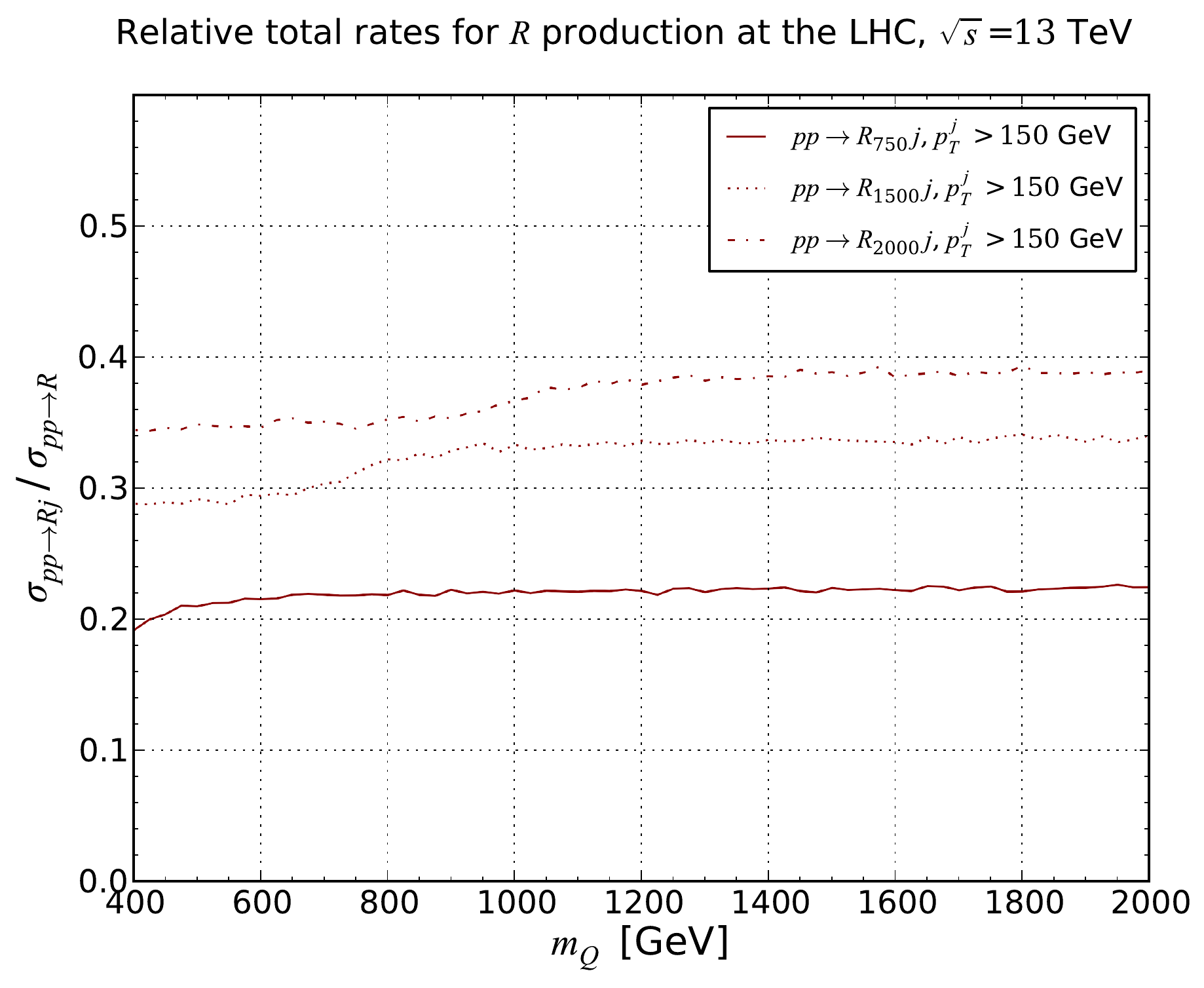}\\
  \caption{\small \label{fig:totalrates2}Total production cross sections for the
  production of a massive scalar resonance, possibly in association with a jet
  whose transverse momentum is imposed to be above a specific threshold. In the
  right panel of the figure, we present the production rates relatively to the
  $p p \to R$ total production cross section. We focus on three scenarios for
  which $m_{\sss R} = 750$~GeV, 1500~GeV and 2000~GeV respectively.}
\end{figure*}

Comparing the \mbox{$pp\to R$} to the \mbox{$pp\to Rj$} predictions, we observe
that any sign for a large beyond the Standard Model contribution to the diphoton
production cross section \mbox{$\sigma(pp\to R \to \gamma \gamma)$} is always
accompanied with significant new physics effects in the production rate of
diphoton events with extra hard jets. In our parameterization of
Eq.~\eqref{eq:LVLQ}, we have considered vector-like quarks lying in the
fundamental representation of the $SU(3)$ group. However, the cross section
ratios (right panel of Figure~\ref{fig:totalrates}) are
independent of the color representation that could be chosen differently,
resulting in larger rates.  The effect of a different color representation
choice for the heavy quark is related to the $SU(3)$ group theory factors of the
different scattering amplitudes,
\begin{eqnarray}
\label{eq:4}
{\cal A}(g^a g^b \to R) &\propto& {\rm Tr}(T^a T^b) = \delta^{ab}/2,\\
\label{eq:5}
{\cal A}_\triangle (g^a g^b \to R g^c) &\propto& \sum_d f^{ab d}{\rm Tr}(T^d T^c)\propto \frac{1}{2}f^{abc}, \\
{\cal A}_\square (g^a g^b \to R g^c) &\propto& {\rm Tr}(T^a [T^b, T^c]) \propto \frac{1}{2} f^{abc},
\label{eq:6}\end{eqnarray}
where $a$, $b$ and $c$ are the color indices carried by the initial-state and
final-state gluons, and ${\cal A}_\triangle$ and ${\cal A}_\square$ are the
one-loop amplitudes related to the \mbox{$gg\to Rg$} process when considering
either the triangle or the box diagrams respectively. The amplitudes are hence
all independent of the specific color representation of the heavy quark.

Investigating the \mbox{$pp\to Rj$} process, it turns out that the full
amplitude exhibits a $t$-channel enhancement such that the contributions from
the triangle diagrams dominate over the box diagram ones, especially when the
mass of the quark $Q$ is not too large. This `triangle dominance' consequently
implies that the
ratio \mbox{$\sigma({pp\to Rj})/\sigma({pp\to R})$} is insensitive to the quark
mass $m_{\sss Q}$ in the mass window of interest, since the same triangle loop
approximately factorizes from the two amplitudes \mbox{${\cal A}(gg\to R)$} and
\mbox{${\cal A}(gg\to Rg) = {\cal A}_\triangle(gg\to Rg) +
{\cal A}_\square(gg\to Rg)$}. The dependence of the
results on the vector-like quark mass $m_{\sss Q}$ therefore stems from the form
of the triangle-loop amplitude ${\cal A}_\triangle ^{(1)}$ that is given, in the
large $m_{\sss Q}$ limit, by
\be
  {\cal A}_\triangle^{(1)}(m_{\sss Q}, m_{\sss R}) \propto \frac{1}{m_{\sss Q}}
   + \frac{7 m_{\sss R}^2}{120 m_{\sss Q}^3}
   + \frac{m_{\sss R}^4}{168 m_{\sss Q}^5}
   + {\cal O}\Big(\frac{m_{\sss R}^6}{m_{\sss Q}^7}\Big)\ .
\label{eq:calA}\ee

On Figure~\ref{fig:totalrates2}, we study the dependence of the previous results
on the mass of the resonance $m_{\sss R}$, and focus on two extra benchmark
scenarios where $m_{\sss R}$ is fixed to 1500~GeV and 2000~GeV respectively. On
the left panel of the figure, we show that the total rate is reduced when
considering heavier $R$ states, as could be expected from the corresponding
phase space suppression and the dependence of ${\cal A}^{(1)}_\triangle$ on
$m_{\sss R}$. We can also observe the presence
of threshold effects related to the imaginary part of the loop-amplitude when
the vector-like quark mass is about half the $R$ scalar mass. Such setups will
nevertheless not be considered in the following, as in this case the scalar
particle preferably decays back into a pair of vector-like quarks (that occurs
at the tree level) and not into a photon pair (that is a loop-induced process).
On the right panel of the figure, we present the dependence of the
\mbox{$\sigma({pp\to Rj})/\sigma({pp\to R})$} ratio in terms of the vector-like
quark mass. We find that the cross section for producing the scalar state in
association with a hard jet is relatively larger and larger for heavier and
heavier $R$ states. This property originates from two contributions, the
renormalization scale choice (and the associated $\alpha_s$ value) that depends
on $m_{\sss R}$ as well as the $m_{\sss R}$ functional form of the triangle
loop-amplitude (see Eq.~\eqref{eq:calA}). The $\alpha_s$ dependence is however reduced when comparing the
two large $R$-mass values.

Similar conclusions could be observed for the new physics parameterization of
Eq.~\eqref{eq:Lsq} as the results of Eq.~\eqref{eq:4},
Eq.~\eqref{eq:5} and Eq.~\eqref{eq:6}
are general enough to hold regardless of the nature of the particle mediating
the $R$ coupling to the Standard Model.

 The loop amplitude is however suppressed by one extra factor
of the mass of the scalar quark running into the loop,
\be
  {\cal A}^{(2)}_\triangle(m_{\sss \tilde q}, m_{\sss R}) \propto \frac{1}{4 m_{\sss \tilde q}^2}
   + \frac{m_{\sss R}^2}{30 m_{\sss \tilde q}^4}
   + \frac{3 m_{\sss R}^4}{560 m_{\sss \tilde q}^6}
   + {\cal O}\Big(\frac{m_{\sss R}^6}{m_{\sss \tilde q}^8}\Big)\ ,
\label{eq:calAsq}\ee
so that it is not possible to simultaneously explain any potential excess of about
10--20~fb in the
diphoton spectrum and maintain the perturbativity of the theory. Scenarios with
extra squarks will therefore not be considered in the rest of this work.

\begin{figure*}
\centering
  \includegraphics[trim={0 0 5.6cm 0},clip,width=.85\columnwidth]{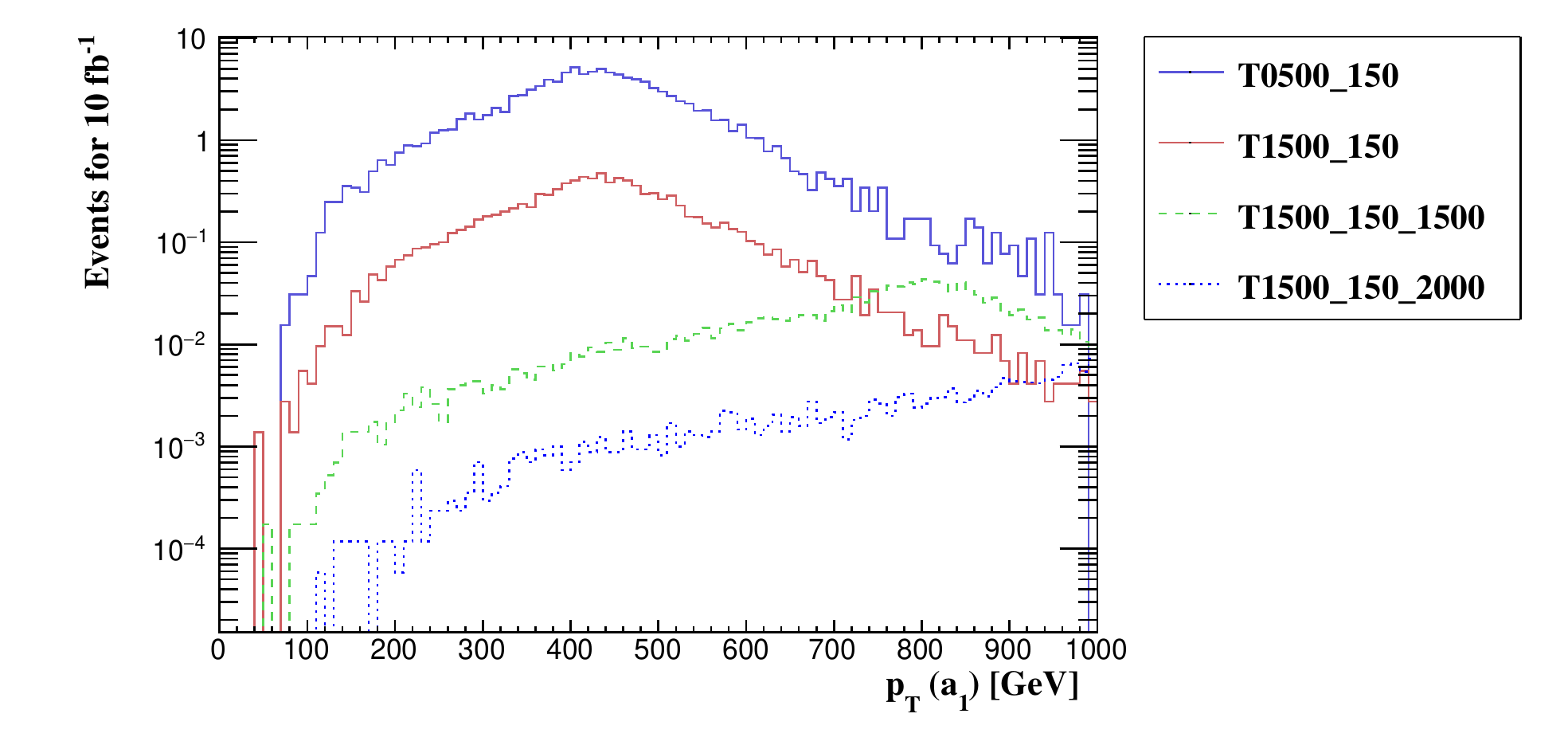}
  \qquad\qquad
  \includegraphics[trim={0 0 5.6cm 0},clip,width=.85\columnwidth]{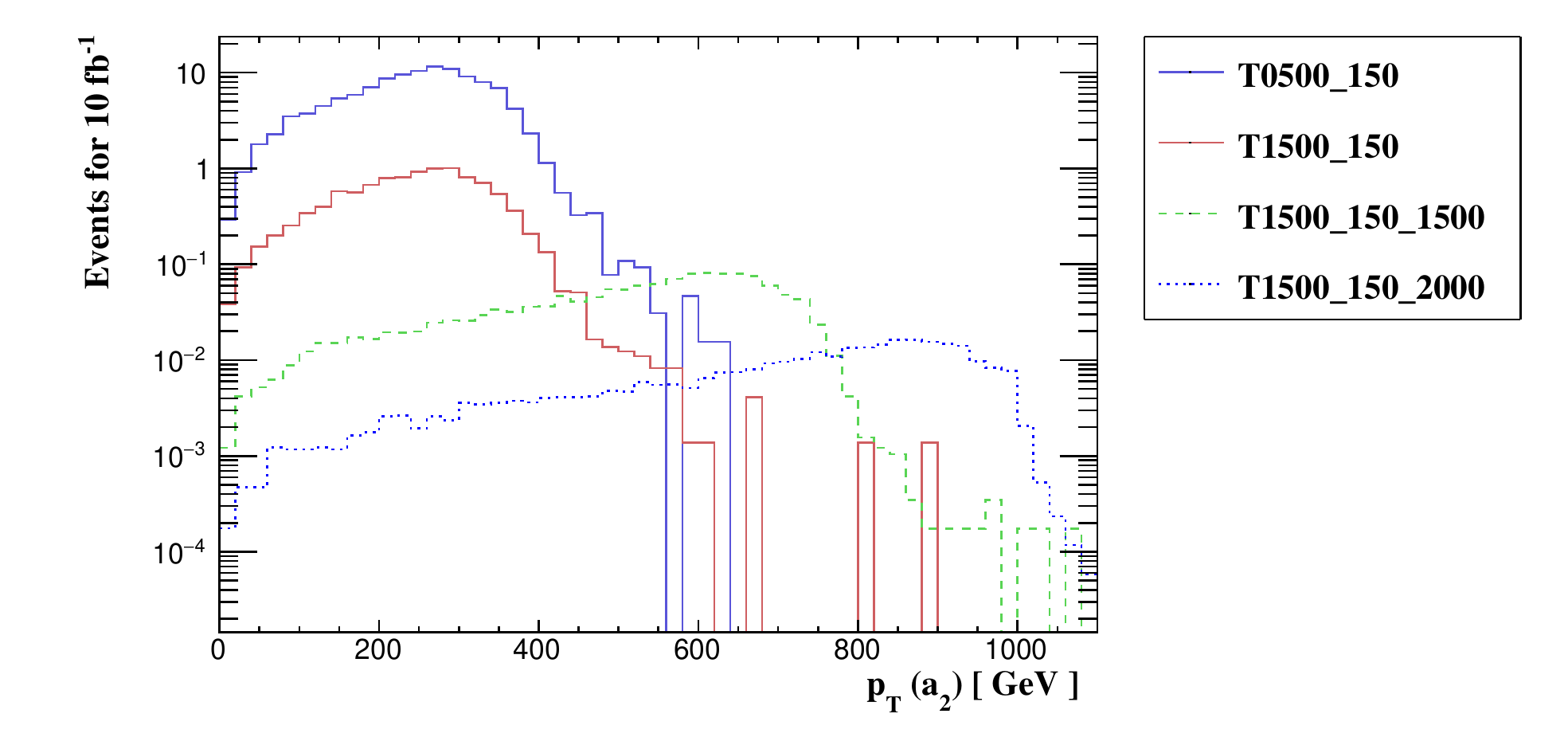}\\
  \includegraphics[trim={0 0 5.6cm 0},clip,width=.85\columnwidth]{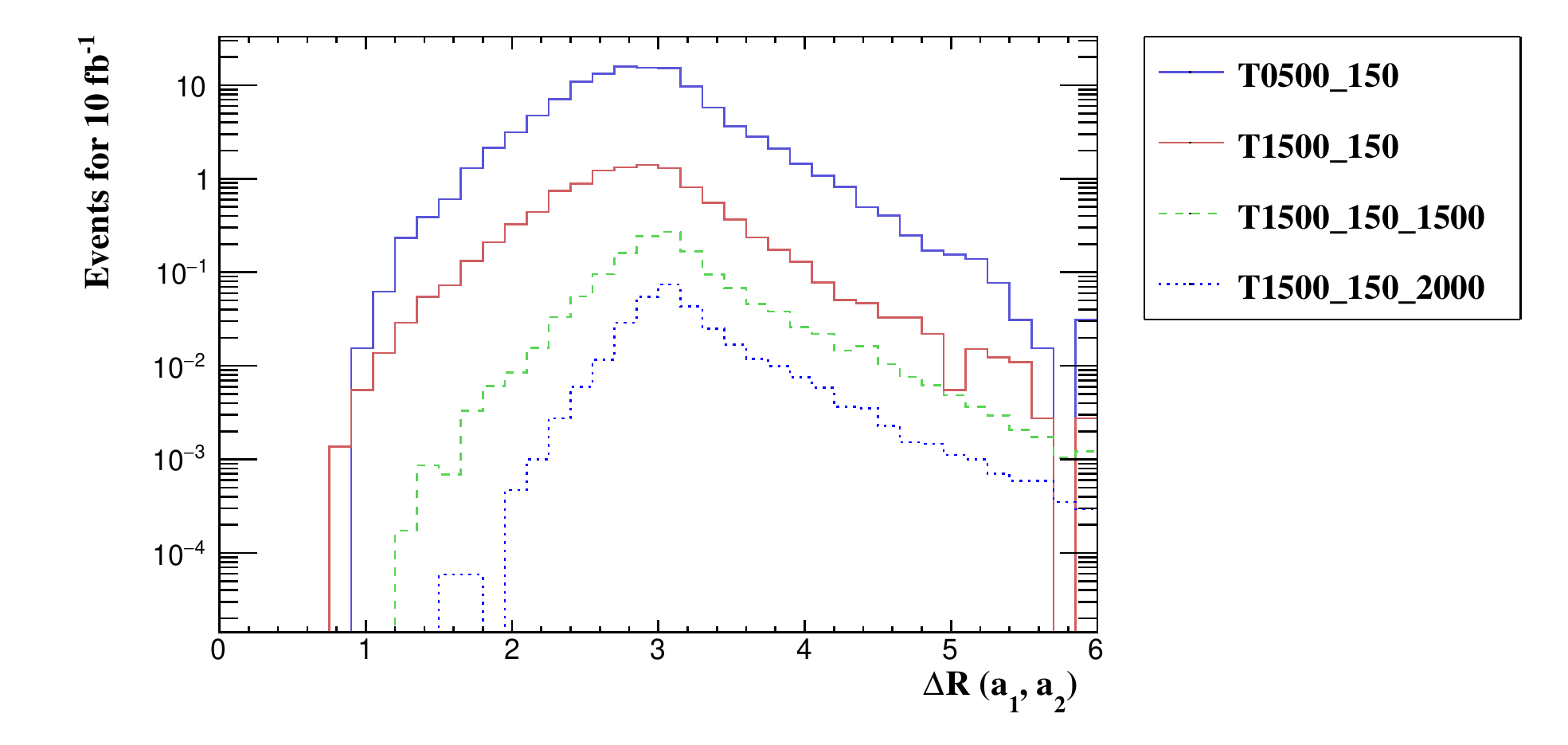}
  \qquad\qquad
  \includegraphics[trim={0 0 5.6cm 0},clip,width=.85\columnwidth]{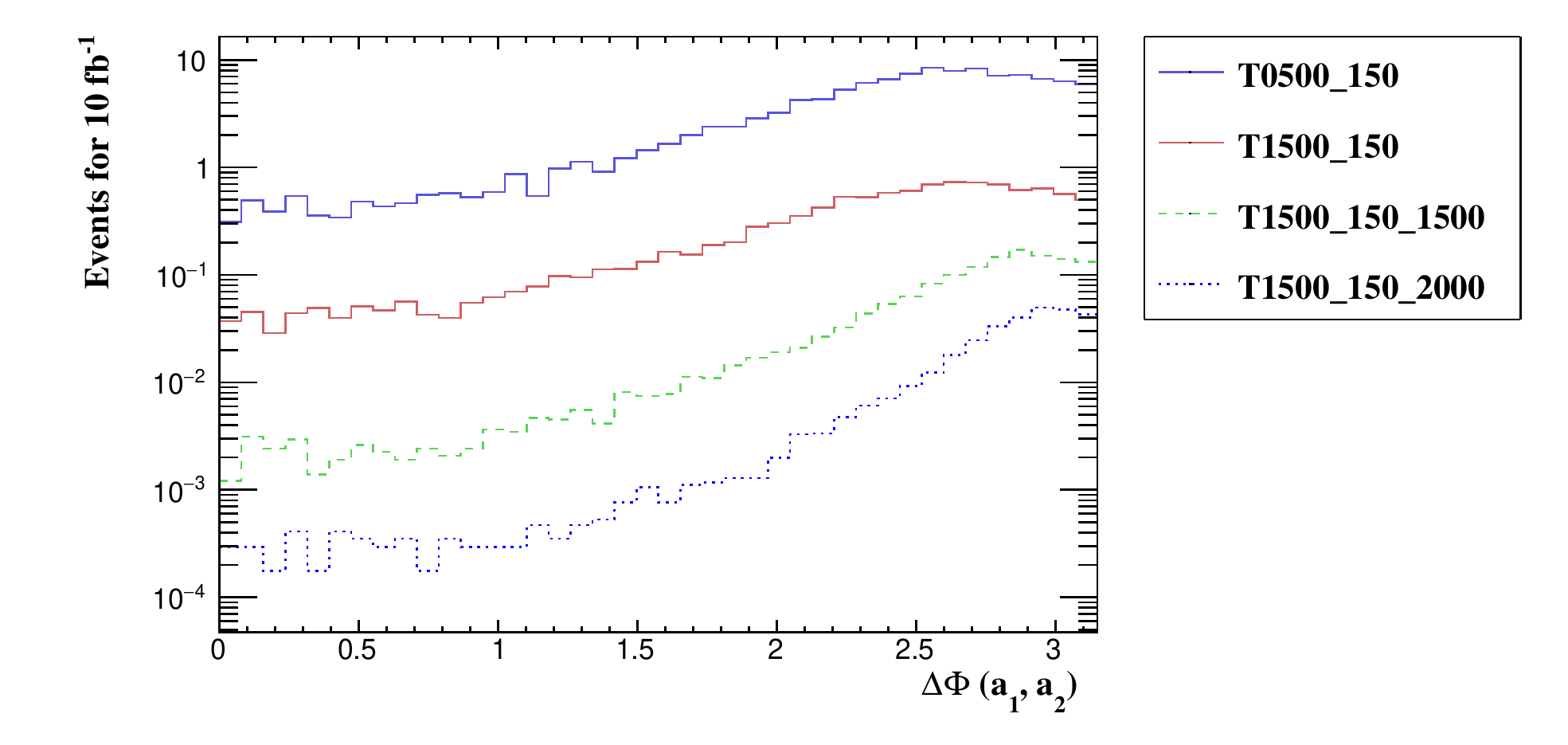}\\
  \caption{\small \label{fig:digamma}Properties of the photon-pair originating
  from the decay of an $R$-particle produced in assocation with a hard jet. We
  present the transverse momentum spectrum of the leading and
  next-to-leading photons (top panel), their angular distance in the transverse
  plane (lower left panel) and their angular distance in azimuth
  (lower right panel). The vector-like quark mass is fixed either to 500~GeV
  (solid
  blue) or the 1500~GeV (solid red) for $m_{\sss R}=$ 750~GeV, and to 1500~GeV
  for scenarios in which $m_{\sss R} =$ 1500~GeV (green dotted) and 2000~GeV
  (purple dotted). The normalization assumes 10~fb$^{-1}$ of 13 TeV LHC
  collisions.}
\end{figure*}

In order to study the properties of a scalar diphoton resonance, it may be
useful to investigate events where the new heavy particle recoils against a hard
jet. For a proper description of such an event configuration, it is necessary to
include at least one extra radiation at the level of the matrix element and
match the fixed-order results to parton showers for a correct modeling of the
remaining jet activity.
We make use of the {\sc MadSpin}~\cite{Artoisenet:2012st} and
{\sc MadWidth}~\cite{Alwall:2014bza} programs to simulate the decay of the
scalar
$R$ particle on the basis of the associated matrix element, after shrinking the
$R\gamma\gamma$ loop-induced interaction to a point-like vertex. We then
interface the partonic events obtained in this way to the parton showering and
hadronization infrastructure implemented in the {\sc Pythia}~8
package~\cite{Sjostrand:2014zea}, and reconstruct all final state jets by means
of the anti-$k_T$ algorithm~\cite{Cacciari:2008gp} with a radius parameter fixed
to 0.4 as implemented in {\sc FastJet}~\cite{Cacciari:2011ma}. We finally
analyze the generated events by means of the {\sc MadAnalysis}~5
platform~\cite{Conte:2012fm}.

On Figure~\ref{fig:digamma}, we study the properties of the diphoton system
originating from the decay of the $R$ particle (when produced in association
with a hard jet)
whose mass has been fixed to $m_{\sss R} = 750$~GeV both in the case
where the heavy quark mass is set to 500~GeV (solid blue curve) and when it is
set to 1500~GeV (solid red line). Our results include
a selection on the leading final-state jet, its transverse momentum $p_T$ being
imposed to be larger than 150~GeV and the absolute value of its pseudorapidity
$|\eta|$ to be smaller than 2.5. On the upper panel of the figure, we present
the transverse momentum distribution of the two final-state photons, and their
angular distances in the transverse plane and in azimuth are shown in the lower
panel of the figure. Except the normalization, the shapes of the spectra are
very similar regardless of the vector-like quark mass choice, and this for all
represented distributions. Although an inclusive diphoton signal is in general
accompanied by a diphoton plus a hard jet signal, it is very unlikely that the
photon properties could help on getting information on the vector-like
quark state
running into the loop. We additionally study two scenarios for which $m_{\sss Q}
 =$ 1500~GeV and $m_{\sss R}$ is respectively fixed to 1.5~TeV (green dotted
line) and 2~TeV (purple dotted line). The conclusions are similar, with the
difference in the shapes of the distributions being driven by the $R$ state mass.

\section{Conclusion \label{sec:conclusion}}

We have studied an extension of the Standard Model containing a singlet scalar
particle $R$ interacting with a new heavy colored quark $Q$ (with a spin
\mbox{$s=1/2$}) or squark $\tilde{q}$ (with a spin \mbox{$s=0$}) through which
the singlet scalar can be produced at the LHC by gluon fusion and decay into a
diphoton system. We have pointed out that the observation of additional jet
activity around an $R$-induced photon pair is important for checking the
consistency of the underlying physics, as significant extra jet production is
also predicted.  

The rate for producing an additional hard jet at the LHC, evaluated relatively
to the $R$ single production cross section has been found to be around
20--40\%, the exact value depending on the mass of the resonance particle
$m_{\sss R}$ that we have varied in the 750--2000~GeV mass window. It is however
less sensitive to the mass of the particle running into the loop amplitude,
$m_{\sss Q}$ and $m_{\sss \tilde{q}}$ for scenarios with a vector-like quark and
scalar quark respectively. The properties of the primary and secondary photons
issued from the $R$ decay have also been investigated, and we have found that
they do not seem to provide additional information on the colored particle
running into the loop other than related to the overall normalization of the
considered distributions, which is expected in the case of a scalar particle
decay. We therefore strongly recommend to correlate the future inclusive
analyses of highly massive diphoton systems at the LHC with less inclusive
analyses imposing strong requirements on the underlying jet activity. 

\acknowledgments
BF has been supported by the {\it Th\'eorie LHC France} initiative of the CNRS
(INP/IN2P3), SC by the National Research Foundation of Korea (NRF) grant funded
by the Korean government (MSIP) (No. 2016R1A2B2016112) and MS by IBS (Project
Code IBS-R018-D1).

\bibliographystyle{JHEP}
\bibliography{biblio}

\end{document}